\documentclass[usenatbib]{basi}
\usepackage[T1]{fontenc}
\usepackage[british]{babel}
\usepackage[varg]{txfonts}
\usepackage{rotating}
\usepackage{dcolumn}

\bibstyle{aasjournal}

\begin{document}
\title[MaStar]{SDSS-IV MaStar: a Large, Comprehensive, and High Quality Empirical Stellar Library}
\author[R. Yan et~al.]%
       {Renbin Yan$^1$\thanks{email: \texttt{yanrenbin@uky.edu}},
       MaStar Team\\
       $^1$Department of Physics and Astronomy, University of Kentucky, 505 Rose Street, Lexington, KY, 40506-0055, USA
}

\pubyear{2012}
\volume{00}
\pagerange{\pageref{firstpage}--\pageref{lastpage}}

\date{Received --- ; accepted ---}

\maketitle
\label{firstpage}

\begin{abstract}
We introduce the ongoing MaStar project, which is going to construct a large, well-calibrated, high quality empirical stellar library with more than 8000 stars covering the wavelength range from 3622 to 10,354A at a resolution of $R\sim2000$, and with better than 3\% relative flux calibration. The spectra are taken using hexagonal fiber bundles feeding the BOSS spectrographs on the 2.5m Sloan Foundation Telescope, by piggybacking on the SDSS-IV/APOGEE-2 observations. Compared to previous efforts of empirical libraries, the MaStar Library will have a more comprehensive stellar parameter coverage, especially in cool dwarfs, low metallicity stars, and stars with different [$\alpha$/Fe]. This is achieved by a target selection method based on large spectroscopic catalogs from APOGEE, LAMOST, and SEGUE, combined with photometric selection. This empirical library will provide a new basis for calibrating theoretical spectral libraries and for stellar population synthesis. In addition, with identical spectral coverage and resolution to the ongoing integral field spectroscopy survey of nearby galaxies --- SDSS-IV/MaNGA (Mapping Nearby Galaxies at APO). this library is ideal for spectral modeling and stellar population analysis of MaNGA data.
\end{abstract}


\section{Motivation}
Stars are the dominant light source in the Universe from ultraviolet to near-infrared. Stellar spectral libraries provide templates for different kinds of stars. Thus, they are essential for the modeling of stellar and galaxy spectra, in nearly all cases where the spectra or spectral energy distributions are involved. 

 Many theoretical and empirical stellar libraries exist, but they have significant shortcomings. Theoretical libraries can have very high spectral resolution, complete wavelength coverage, and can cover areas of parameter space that are not available in the Milky Way. However, they are not yet realistic enough. Many physical effects are difficult to model, such as non-local-thermodynamic-equilibrium, line-blanketing, sphericity, expansion, non-radiative heating, convection, etc. Moreover, an extensive and accurate list of atomic and molecular line opacities is very difficult to construct. 
Thus, theoretical libraries need to be calibrated or corrected with empirical libraries. Current empirical libraries also have several serious shortcomings. The biggest issue with current empirical libraries is their lack of adequate coverage of the parameter space ($T_{\rm eff}$, $\log g$, [Fe/H], [$\alpha$/Fe]). Even within the stellar types and abundance patterns available in the Milky Way, the coverage is quite incomplete. Among current libraries, MILES \citep{Sanchez-Blazquez06} has the most extensive parameter coverage, but it is still far from sufficient in cool dwarfs, TP-AGB stars, metal-poor stars, and hot and young stars. They also do not sufficiently cover the variation in $[{\rm \alpha/Fe}]$ vs. [Fe/H] among stars in the Milky Way. A more extensive coverage in this parameter space is critically needed to understand chemical evolution of galaxies. There is much room for improvement given stars available in the Milky Way. 

\begin{figure}
\centerline{\includegraphics[width=0.9\textwidth, viewport=0 30 730 360, clip]{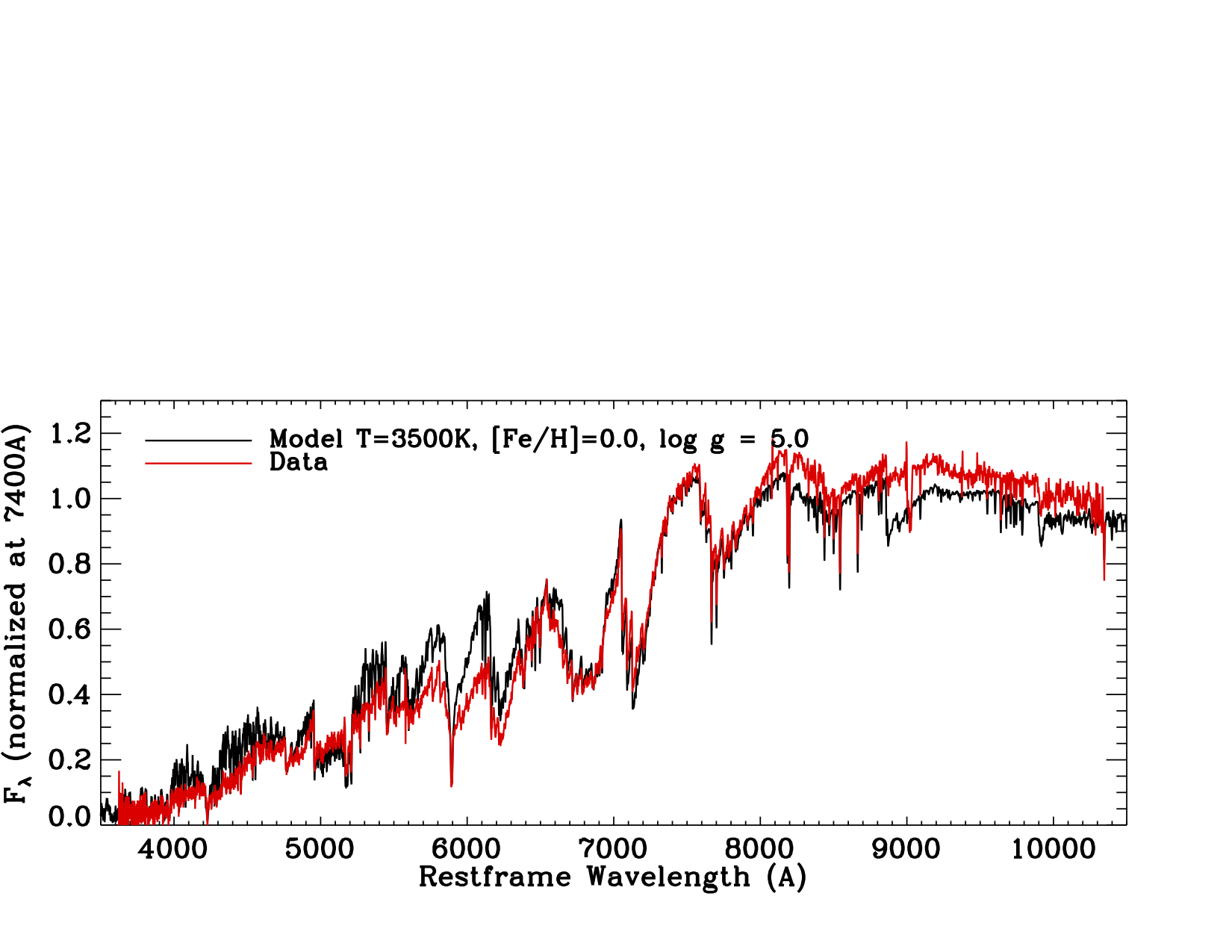}}
\caption{\small An example M-dwarf spectrum from our observations compared to a theoretical model spectrum, based on the ATLAS12/SYNTHE atmosphere and spectrum synthesis codes with the latest atomic and molecular line lists provided by R. Kurucz (See \citealt{ConroyvD12a} for details). The model fits well in some wavelength ranges but fails in others due to an incorrect line list and simplified assumptions in modeling. We need many more well-calibrated empirical spectra to help correct the theoretical models.}
\label{fig:coolstarkuruczcomp}
\end{figure}



Therefore, we are building a new empirical stellar spectral library --- MaNGA Stellar Library or MaStar for short --- to overcome these shortcomings. 


\section{Observations}

MaStar observations are done on the 2.5-meter Sloan Foundation Telescope at the Apache Point Observatory, as an ancillary program of the 4th-generation Sloan Digital Sky Survey (SDSS-IV, \citealt{Blanton17}). SDSS-IV has 3 main survey components: APO Galactic Evolution Experiment-2 (APOGEE-2, \citealt{Majewski16}), Mapping Nearby Galaxies at APO (MaNGA, \citealt{Bundy15,Yan16b}), and the extended Baryon Oscillation Spectroscopic Survey (eBOSS, \citealt{Dawson16}). The use of fibers and plug-plates enables parallel observing between optical and infrared. APOGEE-2's fibers and MaNGA's fiber bundles can be plugged on a plate at the same time with APOGEE fibers feeding the infrared spectrograph and MaNGA fiber bundles feeding the optical spectrographs. Therefore, by piggybacking on APOGEE-2 during bright time, we can make use of the MaNGA fiber bundles to obtain optical spectra of stars. 

Typically, each plate is observed for $4\times15$-minute exposures per night on multiple nights. The number of nights each plate is observed and the cadence is determined by the APOGEE-2 survey. Our stars have magnitudes ranging between 11.7 and 17.5\ in $g$ or $i$ band, yielding a median S/N greater than 100 per resolution elements. The spectra we obtain have exactly the same wavelength coverage and spectral resolution as the MaNGA survey, making our library the ideal set for modeling MaNGA spectra and modeling the stellar population in MaNGA target galaxies. 

\section{Improved Stellar Parameter Coverage}

A major improvement in our library compared to previous efforts is the more comprehensive stellar parameter coverage. We achieved this by selecting targets from existing stellar parameter catalogs wherever possible, such as APOGEE-1 and -2, SEGUE \citep{Lee08a, Lee08b,AllendePrieto08}, and LAMOST \citep{Zhao12,Deng12} stellar parameter catalogs.  
Given the stars available in the planned APOGEE-2 footprint, we compute the local density around each star in the three-dimensional parameter space of $T_{\rm eff}$, $\log g$, and [Fe/H]. We then randomly draw stars with a probability inversely proportional to the local density and obtain a sample of stars that evenly populate the parameter space. 

Not all stars have measurements of [$\alpha$/Fe]. For those stars that do, we adjust the drawing probability among them to balance the sampling in this [$\alpha$/Fe] dimension. 
In addition, we would like to select as many stars as possible from high-fidelity stellar parameter catalogs such as those with higher-resolution spectroscopy. Therefore, we also adjust the drawing probability according to which catalog the stars come from. The probability of APOGEE stars are increased relative to those from SEGUE and LAMOST. 

\begin{figure}[h]
\begin{center}
\vspace{-10pt}
\includegraphics[width=1.0\textwidth]{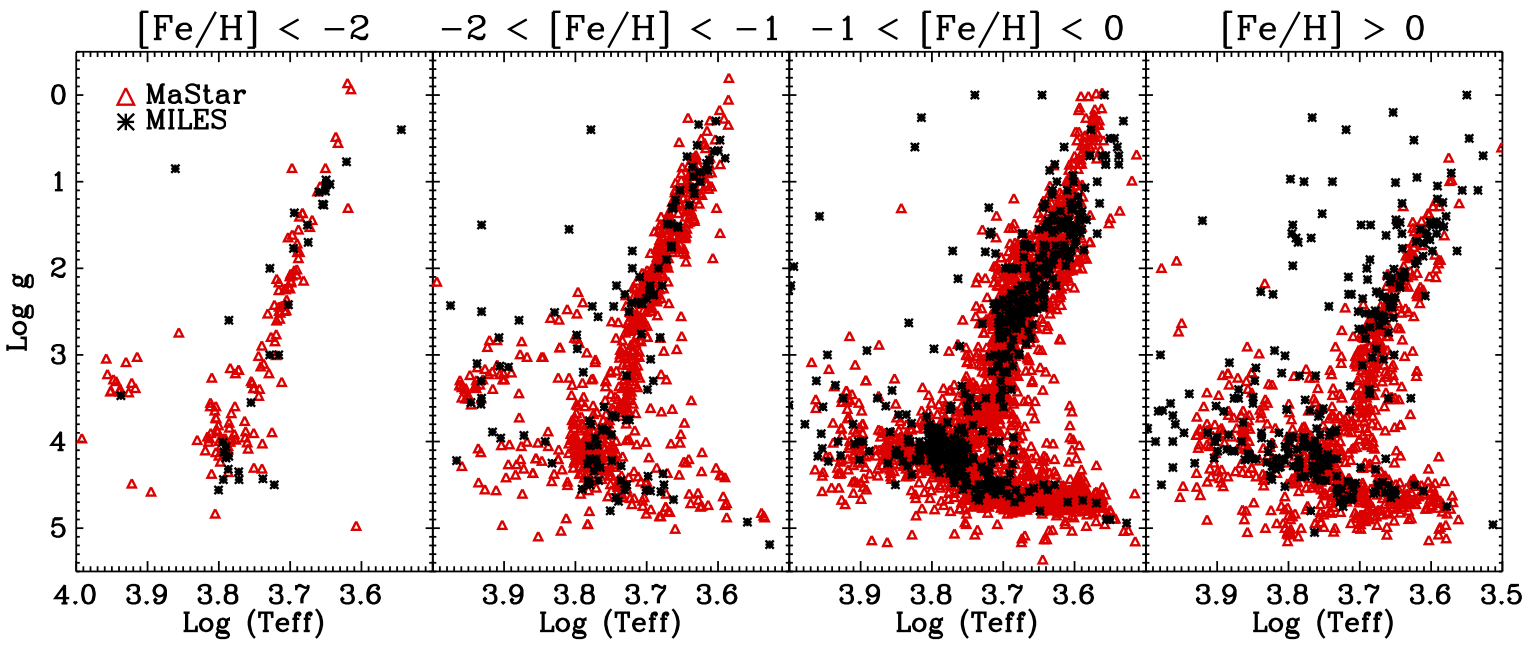}
\vspace{-20pt}
\caption{\small Comparison of the parameter space coverage between the  MILES stellar library and 90\% of all targets in our MaStar library.  In every metallicity bin, we have much denser coverage than MILES, especially in the cool part of the main sequence and the two ends of the giant branch.  {\it Note for our library, the other 10\% of the targets are not shown due to the lack of $\log g$ and [Fe/H] estimates. They are distributed at the hot and cool extremes, providing more extensive coverage than what is shown here.}}
\vspace{-15pt}
\label{fig:coveragecomp}
\end{center}
\end{figure}

Figure~\ref{fig:coveragecomp} shows the expected stellar parameter coverage for 90\% of the targets in MaStar, compared to the parameter coverage of the MILES library. Beside having a much denser sampling, our library also has better coverage at low metallicity and at cool temperatures. For fields without known-parameter stars, we use photometry to estimate the effective temperature and we select very hot and very cool stars to supplement the sample. 

\section{Quality of Spectrophotometry}
The use of fiber bundles instead of single fibers allows us to achieve a much higher accuracy in spectrophotometry calibration.  The calibration is done in a way similar to that for the MaNGA galaxy survey \cite{Yan16}. We observe 12 standard stars simultaneously as the 17 target stars on each plate. The standard stars are chosen to be late F-stars. We compare the observed spectrum with theoretical templates to derive the correction vector. The calibration technique for integral field spectroscopy using fiber bundles is very different from that of slit spectroscopy, and can achieve significantly higher accuracy than single-fiber spectroscopy.  See \cite{Yan16} for details. 

\begin{figure}[h]
\begin{center}
\includegraphics[width=0.8\textwidth,viewport=0 300 770 590, clip, angle=180]{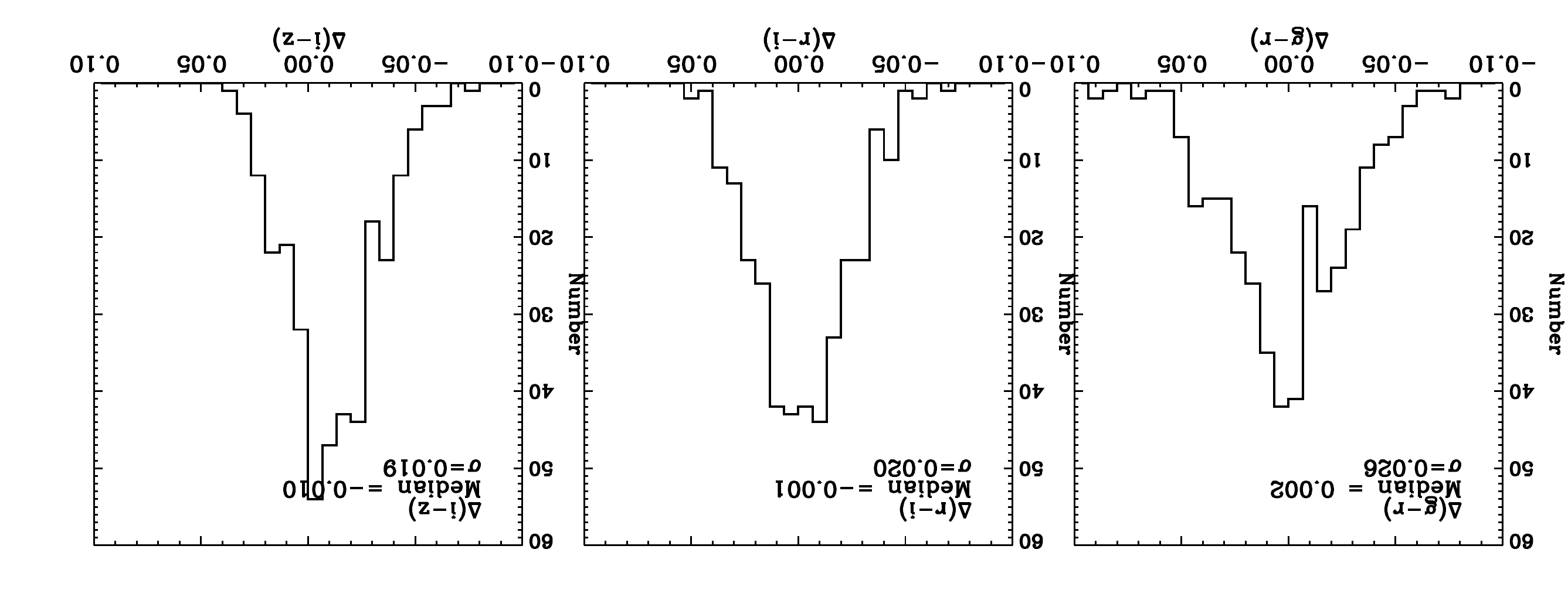}
\vspace{-10pt}
\caption{\small
The distribution of differences between the colors measured from our spectra and those from PanSTARRS-1 photometry. Our relative flux calibration is good to 2-3\% in high galactic latitude fields. For fields at low $|b|$, we are developing a method to accurately correct for extinction to achieve similar accuracy.} 
\vspace{-20pt}
\label{fig:fluxcalaccu}
\end{center}
\end{figure}

Here we illustrate the accuracy of the resulting relative calibration by comparing synthetic colors measured from our spectra to PanSTARRS1 photometry in Fig.~\ref{fig:fluxcalaccu}, demonstrating that we have already achieved a {\it relative} flux calibration accuracy better than 3\% among $griz$ bands.

\section{Summary}

To summarize, we are building a large stellar spectra library, MaStar, with comprehensive stellar parameter coverage and excellent spectrophotometric calibration. The spectra have a wide wavelength coverage (3,622-10,354$\AA$) and a spectral resolution of $R\sim2000$, the same as the MaNGA IFU galaxy survey. The resulting library can be used to model stellar and galaxy spectra directly or be incorporated into stellar population models. 

\bibliographystyle{aasjournal}
\bibliography{astro_refs}
\end{document}